\documentclass[letterpaper,12pt,oneside,onecolumn]{article}%
\usepackage{geometry}%
\usepackage{titlesec}%
\titleformat{\section}[hang]{\normalfont\normalsize\bfseries}{\thesection}{12pt}{\centering}%
\titleformat{\subsection}[display]{\normalfont\normalsize}{\thesubsection}{12pt}{\underline}%
\titleformat{\subsubsection}[runin]{\normalfont\normalsize}{\thesubsubsection}{12pt}{\underline}%
\newcommand{\PaperTitle}[1]{%
\begin{center}%
    \begin{large}%
        \textbf {#1} \\%
    \end{large}%
\end{center}%
}%
\newcommand{\AuthorList}[1]{%
\begin{center}%
    {#1} \\%
\end{center}%
}%
\newcommand{\AuthorAffiliations}[1]{%
\begin{center}%
    {#1} \\%
\end{center}%
}%
\newcommand{\Keywords}[1]{%
\begin{center}%
   Keywords: {#1} \\%
\end{center}%
}%
\usepackage[utf8]{inputenc}
\usepackage{amsmath}
\usepackage{graphicx}
\usepackage{subfigure}
\usepackage[para]{threeparttable}
\begin{document}%
\PaperTitle{STRUCTURE AND THERMODYNAMICAL PROPERTIES OF ZIRCONIUM
HYDRIDES FROM FIRST-PRINCIPLE	}%
%
%
\AuthorList{Jakob Blomqvist (1), Johan Olofsson (2), Anna-Maria Alvarez (2) and Christina Bjerkén (1)}%
\AuthorAffiliations{(1) Div. Material Science, IMP, Malmö University;
Malmö, SWEDEN\\
(2) Materials Technology, Studsvik Nuclear AB; Nyköping, SWEDEN}%
\section{Abstract} Zirconium alloys are used as nuclear fuel cladding material due
to their mechanical and corrosion resistant properties together with their favorable
 cross-section for neutron scattering. At running conditions, however, there will be an increase
of hydrogen in the vicinity of the cladding surface at the water side of the fuel.
The hydrogen will diffuse into the cladding material and at certain conditions, such as 
lower temperatures and external load, hydrides will precipitate out in the material
 and cause well known embrittlement, blistering and other unwanted effects. Using phase-field 
methods it is now possible to model precipitation build-up in metals, for example
as a function of hydrogen concentration, temperature and external load, but the
technique relies on input of parameters, such as the formation energy of the hydrides and matrix. To that
end, we have computed, using the density functional theory (DFT) code GPAW, the latent heat of fusion as well as
solved the crystal structure for three zirconium hydride polymorphs: $\delta$-ZrH$_{1.6}$, $\gamma$-ZrH, and $\epsilon$-ZrH$_{2}$.
\section{Introduction} The presence of hydrogen in certain metals such as Hf-, Ti- and Zr-based alloys can be a major cause of embrittlement due
to precipitation of hydrides (MH$_{x}$) \cite{Coleman2003}. There are at least three mechanisms associated with the hydrogen-
induced mechanical degradation of metals \cite{Myers1992}. These are (1) phase transformations, such as hydride precipitation
induced by combined presence of H and stress, (2) H-enhanced local plasticity, and (3) weakening of grain boundaries
by H. In the following we will take a closer look at the first of these.

Hydrogen embrittlement may occur by repeated hydride precipitation and cleavage on front of a growing crack. The
phenomena involve hydrogen transport and phase transformation (nucleation and growth) near a crack tip which is
affected by the combined effects of stress concentration, temperature distribution and accommodation modes between
hydride and matrix. The situation in Zr-base systems is more complex by the existence of different phases;
the hexagonal close-packed (hcp) low temperature $\alpha$-phase, and the body-centered cubic (bcc) high
temperature $\beta$-phase, which have different susceptibilities to hydrogen embrittlement. The $\alpha$-phase being more
susceptible and having a lower hydrogen solubility and diffusivity than the $\beta$-phase. Alloys containing a mixture of
$\alpha$- and $\beta$-phase usually crack at the $\alpha$/$\beta$-phase interface \cite{Shih1988}.
In presence of flaws, notches or other stress concentrators, the stress-directed movement of hydrogen may also result
in a sub-critical crack growth mechanism, known as delayed hydride cracking (DHC). Under DHC, hydrogen in solid
solution diffuses to the high-stress region in front of the flaw, where it precipitates as brittle hydrides. The hydrided
region grows larger and denser as more hydrogen arrives to the tip of the flaw, until the material is no longer able
to withstand the local tensile stress. Brittle or semi-brittle fracture then ensues, and the flaw grows by an increment
that is comparable in size to the hydrided region ahead of the flaw tip. The process then repeats itself, thus leading
to stepwise crack growth that leaves characteristic striations on the fracture surfaces \cite{Northwood1983}.

It is today possible to model general hydride micro-structure evolution in a primary matrix phase using a phase-field model\cite{Chen2002}.
 For Zr in particular, Ma et al. used a phase-field kinetic model in a two-dimensional setting to investigate the hydride morphology as a
 function of external load\cite{Ma2002, Ma2006}. In any case, the usage of phase-field methods needs input of parameters, such as the latent
 heat of formation, and the accuracy of the model is determined by the source and precision of these parameters. However, only a few of the
 needed parameters that have been published are from reliable experimental sources and some ambiguities still exist regarding, for example,
 the atomic structure, elastic coefficients, and phase-diagram of the H-Zr system. The atomic structure used today are from neutron 
experiments that dates back some 50 years and it is quite debatable whether one should trust them fully. A number of recent works have used
 ab-initio computational methods to try to improve the picture somewhat and several groups have used DFT calculations to present, for example,
 lattice constants and elastic properties for pure Zr as well as some of the known hydride polymorphs\cite{Domain2002, Domain2004, Holliger2009, Zhu, Zhang2010}.
Domain et al. used ultrasoft pseudo potentials (UPP) and plane-waves single electron wave functions together with a
 GGA exchange-correlation (xc) functional PW91 to compute the structural parameters of the hydrides, the H-solution energies in Zr,
 and $\alpha$Zr elastic coefficients\cite{Domain2002}. In a second paper the same group presented DFT-calculations,
 using same methods as their first work, of stacking-fault energies and surface excess energies of solid solution H-$\alpha$-Zr system \cite{Domain2004}.
 Holliger et al. combined DFT-calculations, using UPP and plane-waves,
 similar to Domain's work, with cluster expansion (CE) in order to find the structures of several metastable H-Zr phases\cite{Holliger2009}. Zhu's group presented in a recent study the 
structural as well as the elastic coefficient of the four ZrH$_{x}$ polymorphs also utilizing UPP and xc functional PW91. They, further, presented an analysis of the relative phase-stability
w.r.t temperature for the various hydrides using phonon-calculations evaluated at the gamma-point only\cite{Zhu}. Zhang et al. recently presented electronic 
structure calculation on $\epsilon$-ZrH$_{2}$ using the projector-augmented wave (PAW) method together with xc-functional PBE\cite{Zhang2010}. They presented elastic constants and phonon dispersion and DOS
curves. When comparing verious groups' results it is worth noticing that (a) Zr is notoriusly difficult to model and (b) all of the above cited works,
except the last one, use a pseudo potential approach, which is generally considered less accurate than projector based and all-electron methods, such as PAW, FP-LAPW and FP-LMTO.
 In a recent paper Udagawa et al. even presented an analysis of plane defect in Zr-H systems by computing $\gamma$-surfaces and surface energy
 in these systems using DFT-methods\cite{Udagawa2010}. There still exists some gaps and some cases where published elastic coefficients do not
 match from one group to another. Zhu et al. recently published a list of elastic coefficient of the four ZrH$_{x}$ hydrides but did not note
any other calculations made of these\cite{Zhu}. In the case of $\epsilon$-ZrH$_{2}$, however, at least one other group have published a full 
set of elastic coefficients as well as phonon-based thermodynamical properties, and in the case of some of the elastic coefficients there are
 a clear discrepancy between the groups \cite{Zhang2010}, even though both groups used similar DFT methods. A fourth hydride, called $\zeta$-Zr$_{2}$H,
was recently detected by Zhao et al. \cite{Zhao2008} and will be part of future investigations. In the present study, however, only the most common hydrides ($\delta$-, $\gamma$-, and $\epsilon$-HZr) will be covered.

In light of the importance of the elastic properties to understand stress driven embrittlement and the trouble with which to find experimental data
on pure hydrides it is clear that DFT-calculations is a powerful tool. In order to bring clarity and fill in some blanks it is the goal of this work
 to first present the atomic and crystal structure of the various ZrH-polymorphs as well as the elastic coefficient.
In the papers second part the phonon band structure and the free energies are presented, for the first time for some hydride phases.
Accurate ground state electron structure calculations are essential in order for phonon-based thermodynamical properties to be accurate. From the phonon bands it is possible to compute the Gibb's free energy
of formation as function of temperature and Helmholtz' free energy of formation as a function of pressure. These are necessary in order to properly deduce the phase diagram as well as needed in phase-field
methods that are able to model the micro-evolution of the hydrides, for example as a function of temperature or load.

\section{Calculations}

DFT calculations are performed using the GPAW-code\cite{GPAW}. GPAW implements the PAW formalism using real-space grids to represent
 wave functions, pseudo electron densities and potentials. In these calculations the GGA type Perdew-Burke-Ernzerhof (PBE)
 exchange-correlation functional \cite{PBE}, a real-space grid spacing of h=0.15 Å and a Monkhorst-Pack k-point sampling\cite{MonkhorstPack}
 with 6x6x6 k-points are used.

The grid-spacing and number of k-points used in the calculations was chosen to ensure that the total energy was sufficiently converged with respect to 
parameter of interest.
 Increasing the number of k-points or decreasing the grid-spacing further was seen to change the total energy in the calculations with at most $0.05$ eV.
\begin{figure}
\begin{center}
\subfigure[$\alpha$Zr]{\includegraphics[width=0.2\textwidth]{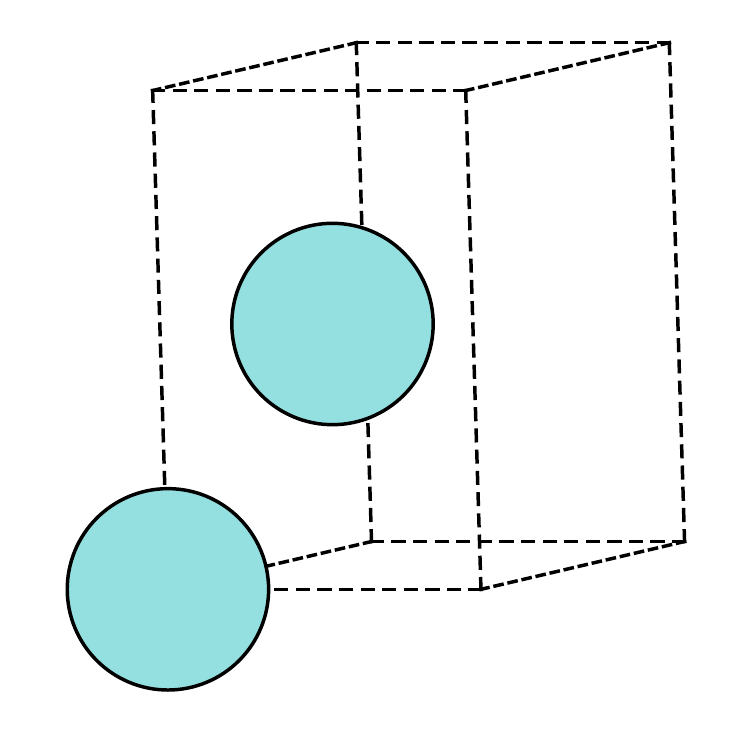}}
\subfigure[$\gamma$-ZrH]{\includegraphics[width=0.2\textwidth]{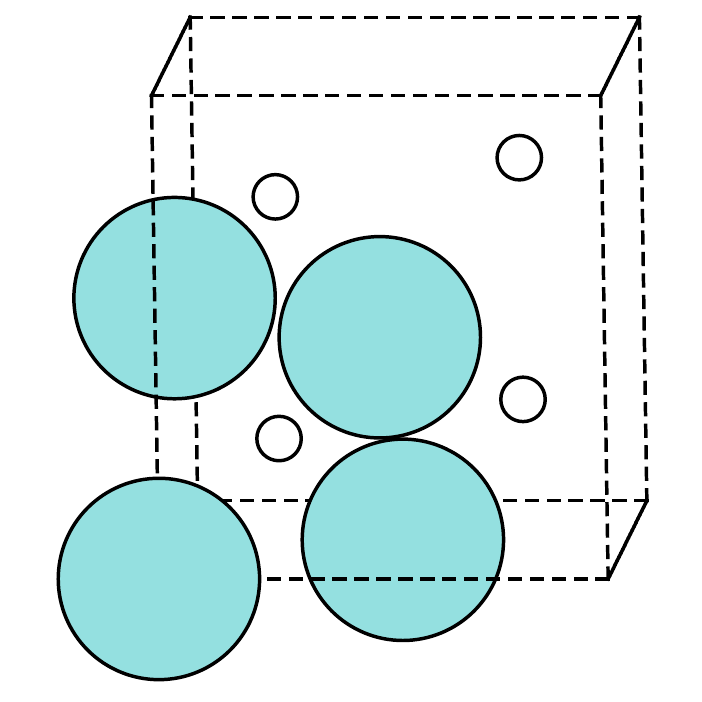}}
\subfigure[$\delta$-ZrH$_{1.5}$]{\includegraphics[width=0.2\textwidth]{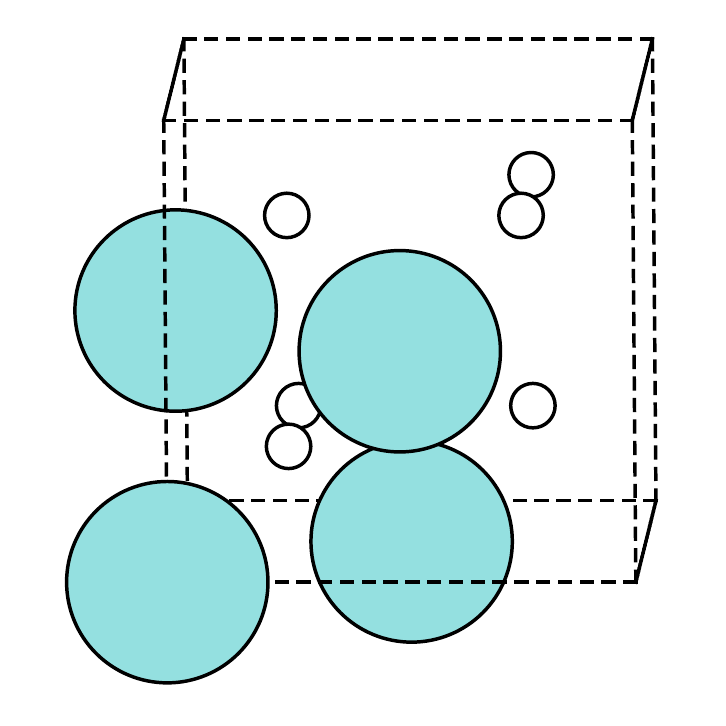}}
\subfigure[$\epsilon$-ZrH$_2$]{\includegraphics[width=0.2\textwidth]{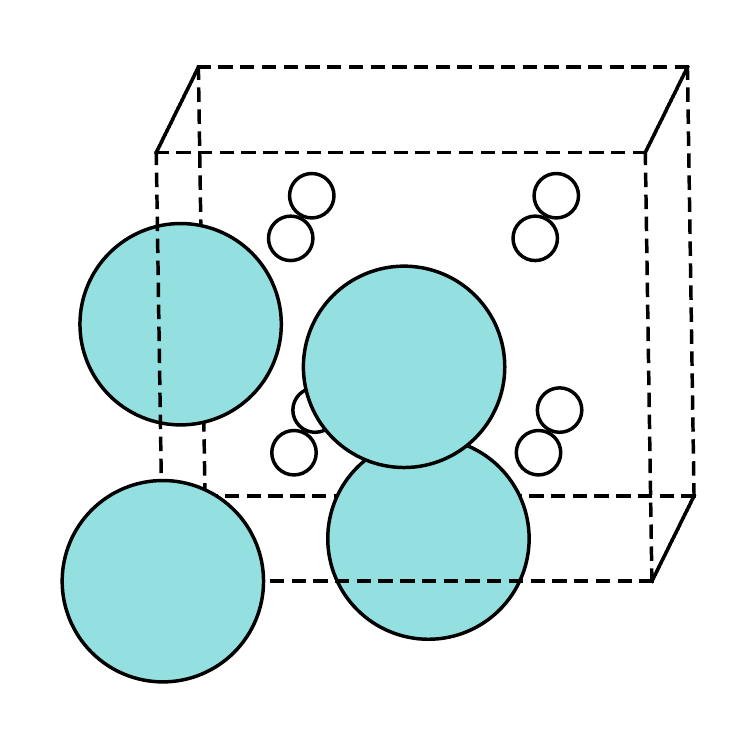}}
\caption{The unit cells for $\alpha$Zr and the different hydrides used in the calculations}
\label{figure_unitCells}
\end{center}
\end{figure}
\subsection{Equilibrium structures}
The equilibrium lattice parameters for $\alpha$-Zr and the different hydrides (see figure \ref{figure_unitCells} are determined by relaxation of ionic positions 
and calculation of the total energy of the compound in question for different values of the lattice parameters. A forth order polynomial is then
 fitted to the data and the minimum of this polynomial corresponds to the equilibrium lattice parameters. The $\delta$-ZrH$_{x}$ is non-stoichiometric
 and experimentally found to have x=1.66. In order to model this phase we use the fcc x=1.5 structure (cf. c in figure \ref{figure_unitCells}).
 $\gamma$-ZrH was modeled using unitcell where hydrogen occupy the four tetragonal sites in the [110] plane. Another possibility exists, where four H occupy
 tetragonal sites in a diamond configuration. It is not included in the results of this work, however, since it has been found to have
a total energy of formation about 0.3 eV higher than that of planar configuration. 
\subsection{Elastic constants}

The elastic constants of the different zirconium hydrides are calculated by deforming the bravais lattice, $\mathbf{R}$, of the crystal to a deformed state, $\mathbf{R'}$ according to:
\begin{equation}
\mathbf{R'} = \mathbf{R}\begin{bmatrix}
1+e_{xx} & \frac{1}{2}e_{xy} & \frac{1}{2}e_{xz} \\
\frac{1}{2}e_{xy} & 1+e_{yy} & \frac{1}{2}e_{yz} \\
\frac{1}{2}e_{xz} & \frac{1}{2}e_{yz} & 1+e_{zz}
\end{bmatrix}
\end{equation}

The change in energy of the crystal due to the deformation can be described by the following formula:
\begin{equation}
U=\frac{E-E_0}{V_0}=\frac{1}{2}\sum_{i=1}^{6}\sum_{j=1}^{6} C_{ij}e_i e_j
\end{equation}
This equation is valid for small strains and the coefficients $C_{ij}$ is the elastic constants of the crystal and $e_i$ and $e_j$ are components of the strain matrix that describes the deformation,
 where the indices $i,j=1,...,6$ corresponds to the different strains, $i,j=1,...,6=xx,yy,zz,yz,xz,yx$.

When the elastic constants for the material is known bulk modulus and shear modulus of the different hydrides can be calculated using standard relationship between B,G and E \cite{Hill1952}.

\subsection{Thermal properties}

The phonon band structure and thermal properties of the different hydrides are calculated using the small
 displacement method as implemented in the phonopy\cite{PhonopyRef} code. In short, the free energy consist of terms involving the atomic energy, the enthalpy and the entropy.
The first is simply the total energy provided by the zero-kelvin DFT calculation. The enthalpy and the entropy both include temperature dependent terms which are functions of the 
phonon states (bulk vibrations). The phonons, in our calculations, was found by the finite displacement method whereby atoms were displaced a small distance.
 For the present study the distance was $0.1$ Å since it was found to be the minimal displacement needed in order to accurately calculate the Hellman-Feynmann forces
 with amplitudes much greater than numerical error noise that would otherwise introduce chaotic behaviour in the results.
 A supercell of $2\times2\times2$ unit cells matrix was used to calculate the phonon band structure for the different hydrides.
 For $\alpha$Zr a supercell with $3\times3\times2$ unit cells and a displacement of $0.1$ Å were used.

\section{Results}

The resulting equilibrium lattice parameters for the different hydrides and $\alpha$-Zr is summarized in table \ref{table_struct} together with experimental data and the results from previous calculations.
 As seen the results from this study is in good agreement with experimental results and also with some of the previous calculations.
\begin{table}
\begin{center}
\begin{threeparttable}
\caption{The resulting equlibrium structures for $\alpha$Zr and the different zirconium hydrides together with experimental data and the results from previous calculations}
\label{table_struct}
\begin{tabular}{c c c c c c c}
\hline\hline
 & \multicolumn{2}{c}{present work}&\multicolumn{2}{c}{experiment}&\multicolumn{2}{c}{previous work} \\
 & a & c & a & c & a & c \\
\hline
$\alpha$-Zr & 3.237 & 5.157 & 3.232\tnote{a} & 5.147\tnote{a} & 3.223\tnote{b} & 5.175\tnote{b} \\
 &  &  & 3.232\tnote{c} & 5.148\tnote{c} & 3.23\tnote{d} & 5.18\tnote{d} \\
$\gamma$-ZrH & 4.592 & 4.998 & 4.596\tnote{c} & 4.969\tnote{c} & 4.586\tnote{b} & 4.9\tnote{b} \\
 &  &  &  &  & 4.58\tnote{d} & 5.04\tnote{d} \\
$\delta$-ZrH$_{1.5}$ & 4.775 & - & 4.771\tnote{a} & - & 4.67\tnote{b} & - \\
 &  &  &  &  & 4.79\tnote{d} & - \\
$\epsilon$-ZrH$_2$ & 4.999 & 4.433 & 4.975\tnote{c} & 4.451\tnote{c} & 4.72\tnote{b} & 4.21\tnote{b} \\
 &  &  & 4.975\tnote{e} & 4.47\tnote{e} & 5.01\tnote{d}& 4.44\tnote{d}  \\
\hline
\end{tabular}
\begin{tablenotes}
\item [a] Reference \cite{Yamanaka2002}, \item [b] Reference \cite{Zhu}, \item [c] Reference \cite{Zuzek}, \item [d] Reference \cite{Domain2002}, \item [e] Reference \cite{Niedzwiedz1993}
\end{tablenotes}
\end{threeparttable}
\end{center}
\end{table}

\begin{table}[htbp]
\begin{center}
\begin{threeparttable}
\caption{Calculated elastic constants ($C_{ij}$), Bulk modulus ($B$), shear modulus ($G$), and Youngs modulus ($E$) for the different hydrides, everything in units of GPa. Comparison with experiments and previous calculations is presented if available}
\label{table_elastic}
\begin{tabular}{c c c c c c c c c c c c c}
\hline\hline
  phase & method& $C_{11}$ & $C_{33}$ & $C_{12}$ & $C_{13}$ & $C_{44}$  & $C_{66}$& $B$ & $G$ &$E$\\
\hline
$\alpha$Zr& This work&157 & 158 & 51 & 62 & 15 & 44 & 91 & 29 & 80\\
 & PW91/UPP\tnote{a}& 142 & 164 &  & 64 & 29& 39 &  92 & & \\
 & PW91/UPP\tnote{c}& 160 & 182 & & 66 & 18 & 51 & 96 & 39 & 103 \\
 & PBE/UPP\tnote{e}& 139 & 163 & 71 & 66 & 26 & & 93 & & \\
 &Expt.\tnote{b}& 155 & 173 & 67 & 65 & 36 & 44 & 97 & &\\
$\gamma$-ZrH&This work& 131 & 176 & 123 & 92  & 64 & 75 & 117 & 32& 88\\
  & PW91/UPP\tnote{c}& 128  & 187 & 126 & 70 &  55  & 66& 117 & 44& 117\\
$\delta$-ZrH$_{1.5}$&This work& 108 & 108& 139 & 139 & 51 & 51 & 128& 24 & 69\\
  & PW91/UPP\tnote{c}& 63 & 65& 28 & 44& 93 &  101  & 47& 63 & 130\\
  &Expt.\tnote{f}& & & & & & & 126 & 50 & 132\\ 
$\epsilon$-ZrH$_2$&This work& 156 & 132& 144 & 107 & 40 & 60&125 & 25 & 73\\
  &PW91/UPP\tnote{c}& 102& 108& 20 & 11& 36& 24& 44 & 37 & 87\\
  &PBE/PAW\tnote{d}& 166&  146& 141 & 107& 31&  61&130 & 29 & 80\\
\hline
\end{tabular}
\begin{tablenotes}
\item [a] Reference \cite{Domain2002}, \item [b] Reference \cite{Fisher1964}, \item [c] Reference \cite{Zhu},\item [d] Reference \cite{Zhang2010},
 \item [e] Reference \cite{Ikehata2004}, \item [f] Reference \cite{Yamanaka2002}
\end{tablenotes}
\end{threeparttable}
\end{center}
\end{table}
\begin{figure}[htbp]
\begin{center}
\includegraphics[width=0.45\textwidth]{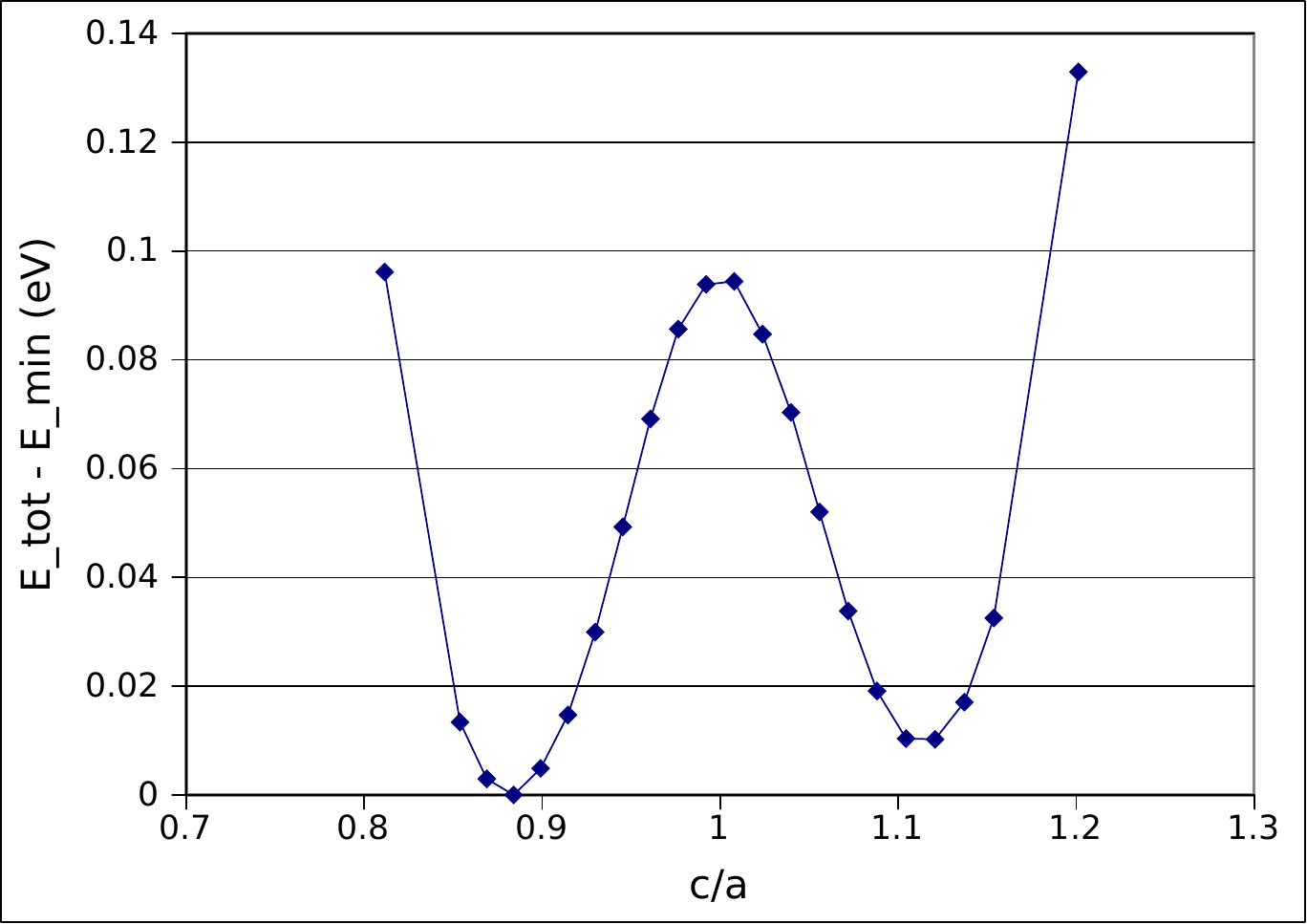}
\caption{Total Energy relative to ground state energy vs. c/a for $\epsilon$-ZrH$_{2}$}
\label{figure_epsilon_E_vs_covera}
\end{center}
\end{figure}
The results from the calculation of the elastic stiffness constants, bulk modulus, shear modulus and Young's modulus for the different
 hydrides are presented in table \ref{table_elastic} together with results from previous calculations and experiments.
The elastic constants of the present work agree nicely with the experimentally determined parameters, where they exist. It is also
clear that for the $\epsilon$-hydride our results agree with those of Zhang et al. \cite{Zhang2010}. As can be seen in figure \ref{figure_epsilon_E_vs_covera}
 we also obtain the ground-state geometry of $c/a < 1$ consistent with experiment. Previous calculations by Ackland\cite{Ackland1998} using Pseudo potential/plane-wave
 methods with PW91 xc-functional found that the lowest energy geometry was when $c/a > 1$. This further shows the difficulty of this system and supports the present results.
 The elastic properties of this hydride do not agree with those presented by Zhu et al. \cite{Zhu} however.
It is not clear to us what is the cause of this discrepancy. Zr is known to be more difficult to calculate accurately than many other elements 
 and since Zhu's group used pseudo-potentials to describe the core-electrons, and plane-waves for the valence electrons, it is possible that the complex
 Zr-H bonds are not described fully enough with their method. Our group and Zhang's, on the other hand, used PAW formalism based methods
 where the full all-electron wavefunction are fully described. Here the projectors are defined either on a grid in k-space or in real space
 in the case of Zhang's group (PAW) and ours (GPAW) respectively. 
\begin{figure}[htbp]
\begin{center}
\subfigure[$\alpha$Zr]{\includegraphics[width=0.45\textwidth]{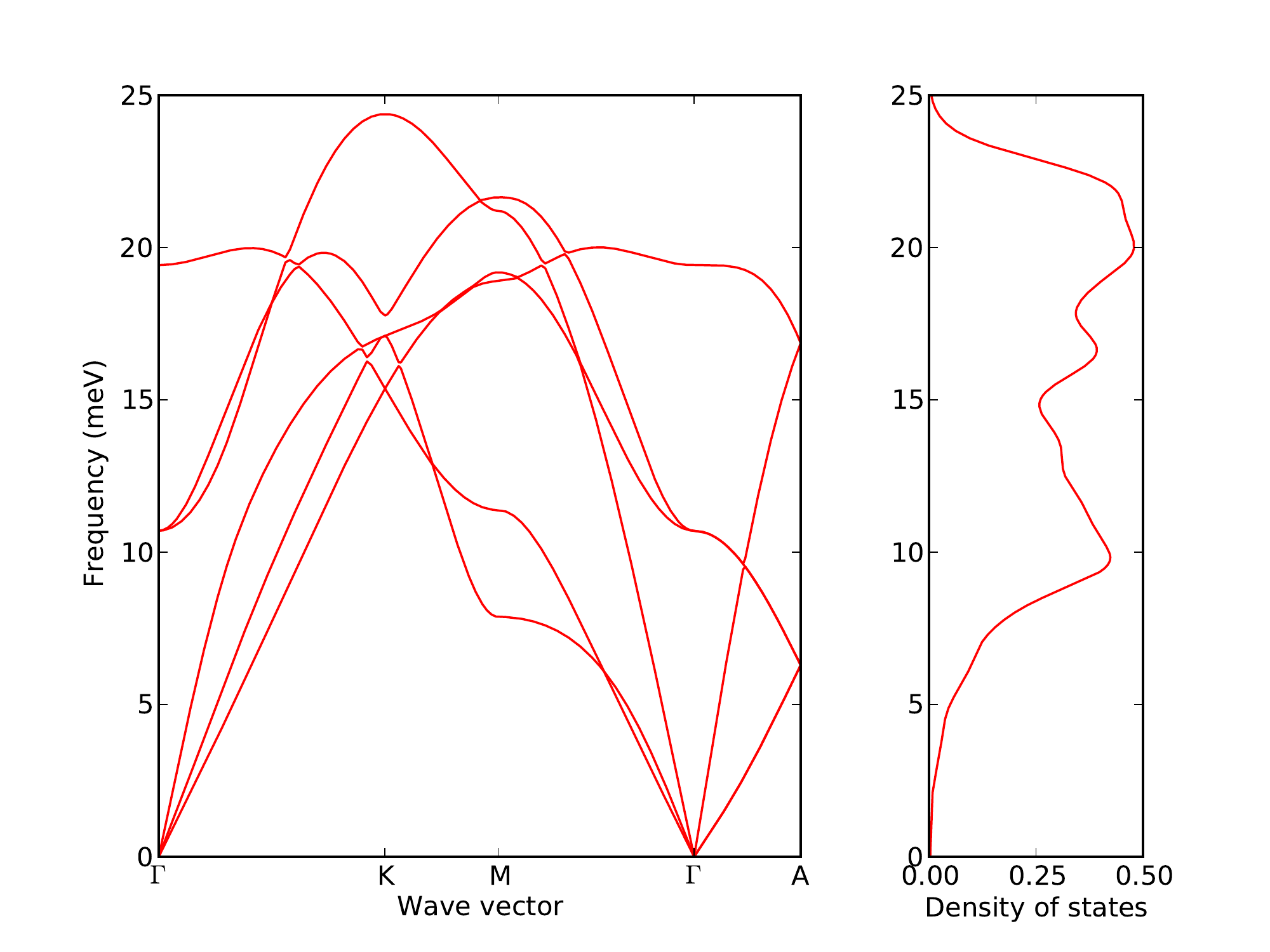}}
\subfigure[$\gamma$-ZrH]{\includegraphics[width=0.45\textwidth]{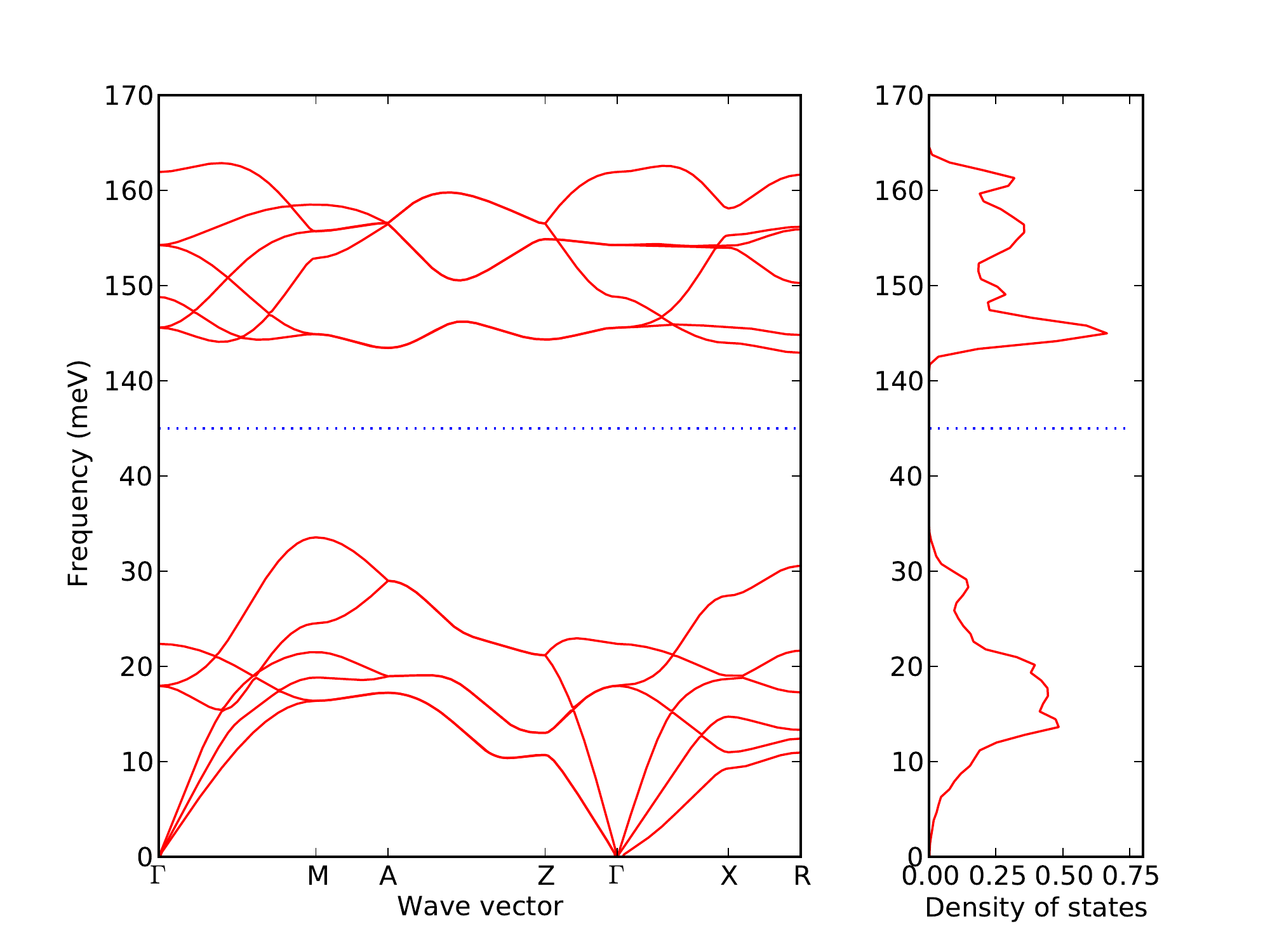}}
\subfigure[$\delta$-ZrH$_{1.5}$]{\includegraphics[width=0.45\textwidth]{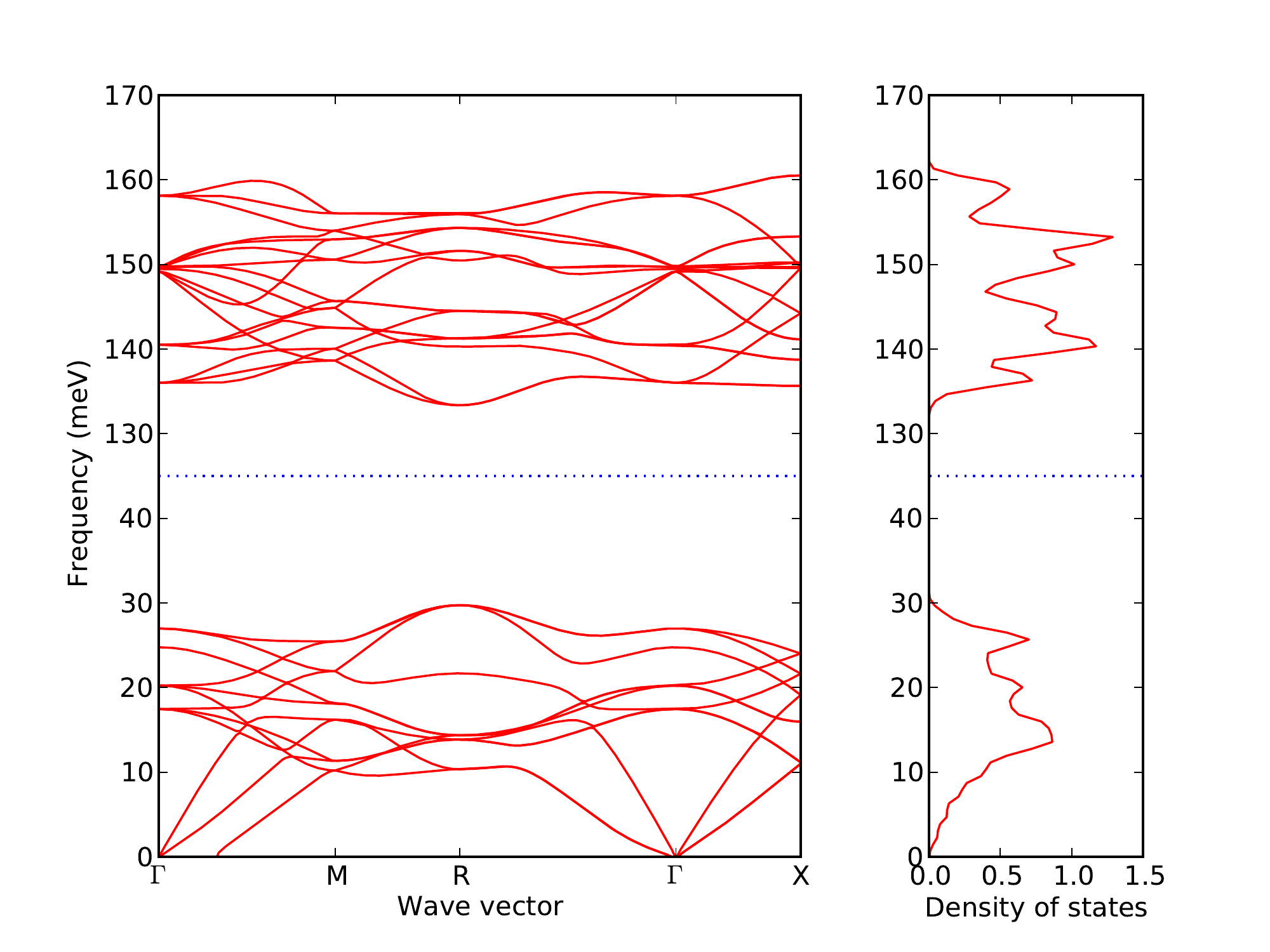}}
\subfigure[$\epsilon$-ZrH$_2$]{\includegraphics[width=0.45\textwidth]{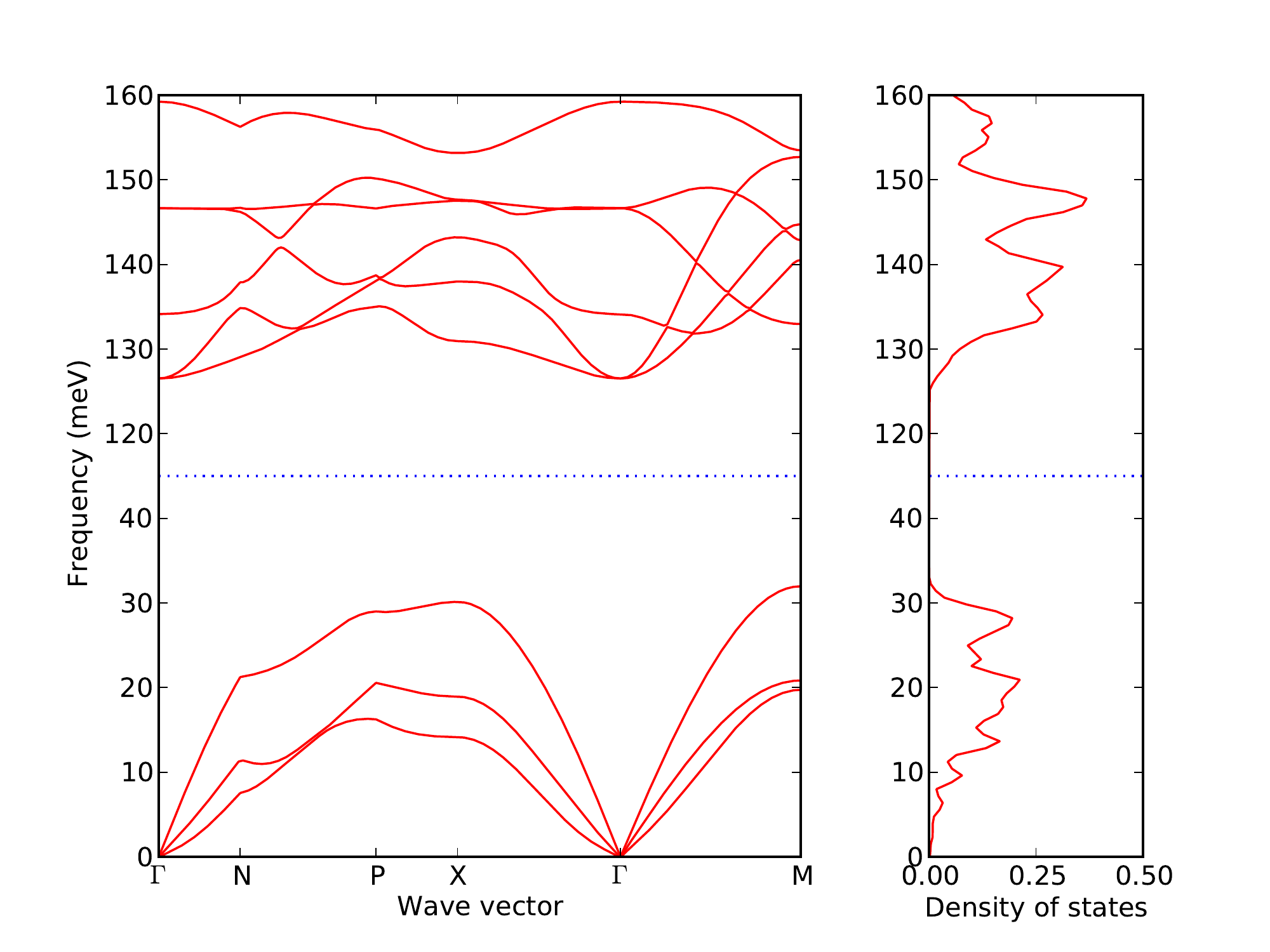}}
\caption{The phonon band structure for $\alpha$Zr and the different hydrides, are frequencies are given in units of meV}
\label{figure_phonon}
\end{center}
\end{figure}

The calculated phonon band structures for $\alpha$Zr and the different hydrides can be seen in figure \ref{figure_phonon}.   
For gamma hydride the unit cell consists of 2 Zr and 2 H and for the delta hydride the phonon dispersion was calculated using a unit cell of 
4 Zr and 6 H. The other hydrides and the hcp-Zr were calculated with primitive unit cells. The band structure for $\alpha$Zr has
 recently been calculated by Souvatzis and Eriksson\cite{Souvatzis2008} and their results are in good agreement with our calculations. The
 phonon band structure for $\epsilon$-ZrH$_2$ was calculated by Zhang et al \cite{Zhang2010} and their results are in qualitative agreement with ours. This is
is further support of the present results.

In figure \ref{figure_phonon} the phonon density of states for $\alpha$Zr and the different hydrides can also be seen to the right of the phonon dispersion.
 In these figures we clearly see that there is a large gap in the phonon band structure for the different hydrides. We have low frequency vibrations from the heavy zirconium atoms and high energy vibrations from the hydrogen atoms.
We also calculated the free energy of formation of the different hydrides, in table \ref{table_free_energy} the free energy of formation for the different hydrides at room temperature and $625$ K ($~350^\circ$ C). From this table we can see the linear decrease of
formation energy for every new H-Zr bond. Of course, this does not imply that $\epsilon$-ZrH$_2$ always have lowest formation energy, and thereby would always be preferred by nature
relative the other hydrides; only that at hydrogen concentration that is stoichiometric relative to the particular hydride these values holds. It simply follows from the lower energy the system feel per added new Zr-H bond.
 To compare relative thermodynamical stability at a general hydride concentration
one needs to formulate formation equation for each hydride in relation to a given hydride and then compute new formation energies. This will be done in future studies.

\begin{table}[htbp]
\caption{The Free energy of formation (in eV) for the three different hydrides at room temperature and $625$ K ($~350^\circ$ C) and at zero pressure.}
\centering
\label{table_free_energy}
\begin{tabular}{c c c}
\hline \hline
 & $295$ K& $625$ K \\
\hline
$\gamma$-ZrH & -0.6002 & -0.3528\\ 
$\delta$-ZrH$_{1.5}$ & -0.8820 & -0.5240 \\
$\epsilon$-ZrH$_{2}$ & -1.2101 & -0.7214 \\ \hline
\end{tabular}
\end{table}
\section{Conclusion}
The structural and thermodynamical properties of the three most common zirconium hydride polymorphs have been presented. Based on a novel all-electron
 dft-method using PAW formalism in evaluated using a real space grid sampling we have presented lattice parameters, elastic stiffness constants at zero pressure and temperature.
In order to learn more about the finite temperature effect the phonon dispersion and DOS was calculated. Thermodynamical properties at two temperatures where finaly presented.
\section{Acknowledgements}
This investigation was supported by grants from Swedish Knowledge Foundation (KKS-project 2008/0503)
\bibliographystyle{unsrt}
\bibliography{ref.bib}

\end{document}